\documentclass[preprint,pre,showpacs,showkeys]{revtex4}%
\usepackage{amssymb}
\usepackage{amsmath}
\usepackage{amsfonts}
\usepackage{graphicx}%

\providecommand{\abs}[1]{\left| #1 \right|} %
\providecommand{\av}[1]{\left\langle #1 \right\rangle} %
\providecommand{\rbr}[1]{\left( #1 \right)}%
\providecommand{\refe}[1]{(\ref{#1})} %

\def\NeG{\mathrm{G}}
\def\BG{\mathrm{BG}}
\def\Ts{\mathrm{T}}
\def\Re{\mathrm{R}}

\begin{document}

\title[ ]{Properties of the Nonextensive Gaussian entropy}
\author{Thomas Oikonomou}
\email{thoikonomou@chem.demokritos.gr}
\affiliation{Institute of Physical Chemistry, National Center for Scientific Research ``Demokritos", 15310 Athens, Greece\\ \\School of Medicine
Department of Biological Chemistry, University of Athens, Goudi, 11527 Athens, Greece}

\begin{abstract}
The present work investigates the Lesche stability (experimental robustness), the thermodynamic stability, the Legendre structure of
thermodynamics, and derives the Maximum Entropy distribution of the one--parametric ``nonextensive Gaussian'' entropy. We show that this entropy
definition fulfills both stability conditions for all values of its parameter ($q\in\mathbb{R}$). The entropy maximizer contains the Lambert
$W$--function, which allows the preservation of the Legendre transformations.
\end{abstract}
\eid{ }
\date{\today }
\startpage{1}
\endpage{1}
\maketitle

\section{Introduction}
A very important quantity in statistical mechanics is the entropy. In the last years there is a great effort in this field to generalize the
concept of thermal equilibrium entropy, which is the Boltzmann--Gibbs (BG) entropy, given by
\begin{equation}
\label{eq:0} S_{\BG}=\sum_{i=1}^{W}p_i \ln{\rbr{1/p_i}}.
\end{equation}
$W$ is the total number of the accessible microstates and $p_i$ their associated probabilities. A variety of systems whose behavior can not be
sufficiently described by the BG statistics caused the tendency in this direction. The candidate entropies can be categorized into two classes,
the trace-form and the non-trace-form ones. The earlier present the structure $\sum_{i}p_i\Lambda(p_i)$, where $\Lambda(p_i)$ in general can be
an arbitrary function, while the latter do not. The entropy that dominates as a possible generalization of the BG entropy in the first class is
the nonextensive Tsallis \cite{Tsallis1988} entropy defined as
\begin{equation}
\label{eq:2} S_{q}^{\Ts}=\frac{1}{1-q}\left( \sum \limits_{i=1}^{W}p_{i}^{q}-1\right),
\end{equation}
and in the second class is the R\'enyi \cite{Renyi1970} entropy defined as
\begin{equation}
\label{eq:1} S_{q}^{\Re}=\frac{1}{1-q}\ln \left( \sum \limits_{i=1}^{W}p_{i}^{q}\right).
\end{equation}
Both definitions in the limit $q=1$ tend to Eq. (\ref{eq:0}). In contrast to the BG entropy, whose maximization, according to the Maximum
Entropy (MaxEnt) principle, leads to exponential distributions, the two latter entropy definitions may also lead to power-law distributions.
With respect to the information theory Eq. (\ref{eq:2}) and Eq. (\ref{eq:1}) can be considered as a limit of the two--parametric Sharma--Mittal
(SM) entropy \cite{SharmaMittal1975}, which is given by
\begin{equation}
\label{eq:3} S_{\left\{ r,q\right\} }^{\mathrm{SM}}=\frac{1}{1-q}\left( {\left[ \sum \limits_{i=1}^{W}p_{i}^{r}\right]
}^{\frac{1-q}{1-r}}-1\right).
\end{equation}
From Eq. (\ref{eq:3}) one can see that for $r=q$ the SM entropy reduces to the Tsallis one, and for $q=1$ this reduces to the R\'enyi one.

However, for $r=1$ the Sharma--Mittal entropy contains yet a third choice of a generalized entropy, which combines the nonextensivity of Tsallis
and the non-trace-form of R\'enyi entropy
\begin{equation}
\label{eq:4} S_{\left\{ r\rightarrow 1,q\right\} }^{\mathrm{SM}} =\frac{1}{1-q}\left( \prod \limits_{i=1}^{W}p_{i}^{(q-1)p_{i}}-1\right)
=\frac{1}{1-q}\Big( e^{\left( 1-q\right) S_{\BG}}-1\Big) =:S_{q}^{\NeG},
\end{equation}
The authors in Ref. \cite{FrankDaff2000} suggested to call Eq. (\ref{eq:4}) ``nonextensive Gaussian'' (NeG) entropy. One of these authors,
Frank, in Ref. \cite{Frank2004} used this definition to solve a nonlinear Fokker--Planck equation which describes an Ornstein--Uhlenbeck
process, in order to obtain analytical expressions for the transition probability densities. In the present article we shall keep this
denomination.

We can easily see that the main difference of the nonextensive Gaussian entropy to the R\'enyi and Tsallis entropy is that the probability
functional is a product of a combination of probabilities instead of sum and hereafter clearly distinguishes from the trace-form entropies.
Another point is that the entire probability functional is raised to a power, in contrast to the probability functionals in Eq. (\ref{eq:2}) and
Eq. (\ref{eq:1}) where every state-probability separately is raised to a power. Written in a different way one can observe that NeG entropy
contains the entire structure of the BG entropy.

The purpose of the current work is to explore some statistical properties of the nonextensive Gaussian entropy and to present the connection to
thermodynamics, in order to complete the study of all three limits, of the Sharma--Mittal definition, which lead to independent entropy
structures. We show that Eq. (\ref{eq:4}), satisfies the nonextensive thermodynamic stability condition and the Lesche stability criterion. The
fulfillment of both conditions is valid for all $q\in \mathbb{R}$. Thus it could be also a good alternative representative of generalized
non-trace-form entropies in statistical thermodynamics. We also derive the maximum-entropy probability for $S_q^{\NeG}$ and present the
connection to the thermodynamical structure.

One can obtain the Tsallis entropy in a heuristic way by replacing the logarithm in the BG entropy with a generalized one. Here we present a
heuristic way to obtain the entropy in Eq. (\ref{eq:4}) by generalizing also the BG entropy using actually the same generalized logarithmic
function. In an integral form the BG entropy can be written as
\begin{equation}
\label{eq:5} S_{\BG}=\int _{1}^{f( p_{i}) }\frac{dx}{x}, \qquad\mathrm{with}\qquad f(p_{i}) =\exp \left[ \sum
\limits_{i=1}^{W}p_{i}\ln{\rbr{1/p_{i}}}\right].
\end{equation}
There are two possible ways to generalize the BG entropy using the concept of a generalized logarithmic function  (see Section \ref{sec5}). The
one leads to the Tsallis entropy and the other leads to the NeG entropy:
\begin{align}\label{eq:6}
\begin{aligned}
S_{q}^{\Ts}&=\int _{1}^{f_q( p_{i}) }\frac{dx}{x}, \qquad\mathrm{with}\qquad
f_q(p_{i}) =\exp \left[ \sum \limits_{i=1}^{W}p_{i}\ln_q{\rbr{1/p_i}}\right],\\
S_{q}^{\NeG}&=\int _{1}^{f( p_{i}) }\frac{dx}{x^{q}}, \qquad\mathrm{with}\qquad f(p_{i}) =\exp \left[ \sum
\limits_{i=1}^{W}p_{i}\ln{\rbr{1/p_i}}\right].
\end{aligned}
\end{align}
For equal probabilities $p_{i}=1/W$, both $S_{q}^{\NeG}$ and $S_{q}^{\Ts}$ tend to
\begin{equation}
\label{eq:7} S_{q}^{\NeG}=S_{q}^{\Ts}=\ln_{q}(W):= \frac{W^{1-q}-1}{1-q} \qquad\left(\ln_1(W)=\ln(W)\right),
\end{equation}
which is the generalized logarithm introduced by Tsallis and coworkers \cite{TsallisMendesPlastino}. One can verify that the additivity rule for
$S_{q}^{\NeG}$ is the same as in the case of Tsallis entropy, namely
\begin{equation}
\label{eq:8} S_{q}^{\NeG}(A+B) =S_{q}^{\NeG}( A) +S_{q}^{\NeG}( B) +\left( 1-q\right) S_{q}^{\NeG}( A) S_{q}^{\NeG}( B).
\end{equation}

Under the concept of stability of a state functional \cite{Lesche2004}, or ``experimental robustness'' as Tsallis proposed in Ref.
\cite{TsallisBrigatti2004} in order to avoid confusions with the thermodynamic stability (see Section \ref{sec3}), we understand the following:
by making a measure, we obtain a distribution of probabilities ${\{p_{i}\}}_{i=1, 2, \cdots, W}$. Repeating the same experiment we obtain a new
distribution of probabilities ${\{{p'}_{i}\}}_{i=1,2, \cdots, W}$ which may be slightly different from the previous one. Now, if we use a
statistical entropy $S$, then we expect that its value should not change dramatically for these two slightly different distributions
${\{p_{i}\}}_{i=1, 2, \cdots, W}$ and ${\{{p'}_{i}\}}_{i=1, 2, \cdots, W}$. Then the entropy $S$ is stable or experimentally robust and is of
physical relevance. The essence of this kind of stability lies in the existence of an entropy associated observable. Lesche in 1982
\cite{Lesche1982} formulated a condition (Lesche stability), which reflects the property of experimental robustness, as follows:
\begin{equation}
\label{eq:9} \rbr{\forall \varepsilon >0} \ \ \rbr{\exists \delta_{\varepsilon} >0} \ \ \rbr{ \left\| p-p'\right\|_{1} < \delta _{\varepsilon }
\Rightarrow \abs{\frac{S(p) -S(p') }{S_{\max}}} < \varepsilon},
\end{equation}
for any value of $W$, where ${\|A\|}_{1}=\sum_{i=1}^{W}|A|$ and $S_{\max }$ is the maximum value of $S$. In the same work he showed that Shannon
entropy is stable, while the R\'enyi entropy does not fulfill this condition for $q\neq 1$. After this, several entropy definitions have been
explored with regard to this criterion. Some of these, which passed this test, are the Abe entropy \cite{AbeKanScar2004}, Tsallis entropy
\cite{Abe2002} and Kaniadakis $\kappa $--entropy \cite{KaniadakisScarfone2004}. Common property of these entropies is their trace-form state
functional. It is striking that non of the non-trace-form examined entropies like R\'enyi, Landsberg-Vedral and escort entropy fulfills the
Lesche criterion.

Another very important condition that an entropy definition has to satisfy, is the thermodynamic stability condition, which is equivalent to the
positivity of the heat capacity $C:= \partial U/\partial T=[-T^{2}\{\partial ^{2}S_{\BG}/\partial U^{2}\}]^{-1}$. $U$ is the internal energy and
$T:= {(\partial S_{\BG}/\partial U)}^{-1}$ the temperature. It is well known that in the case of BG entropy the thermodynamical stability
condition (TSC) and concavity are equivalent to each other:
\begin{equation}
\label{eq:10} \frac{\partial^{2}S_{\BG}( U) }{\partial U^{2}}\leqslant 0 .
\end{equation}
The physical background of Eq. (\ref{eq:10}) is based on the combination of the entropy maximum principle and the additivity of (\ref{eq:0}).
Thus the fulfillment of the condition (\ref{eq:10}) is a very important point in statistical thermodynamics. However, the demand of the
concavity for a nonextensive entropy definition does not suffice to preserve the thermodynamic stability \cite{Wada2004,ScarfoneWada2006}. This
can be easily understood since the total entropy of two composed subsystems is not the sum of the partial entropies of each subsystem.

In Section \ref{sec2} we present the proof of the fulfillment of the Lesche stability criterion. In Section \ref{sec3} we derive in two
different ways the thermodynamic stability condition for the nonextensive Gaussian entropy and show that this is satisfied by the latter. In
Section \ref{sec4} we present the connection of $S_q^{\NeG}$ to thermodynamics. In the final section we draw our main conclusions.

\section{Lesche stability}
\label{sec2}

The Lesche stability condition reflects the reproducibility of the values of any observable quantity. Here we prove this condition with regard
to $S_{q}^{\NeG}$. The exponential form of the $S_{q}^{\NeG}$--state functional makes difficult, if not impossible, the usage of the formalism
in Ref. \cite{Abe2002}. In order to overcome this problem we shall try to find a probability functional which is greater than the expression
$\exp [ (1-q)S_{\BG}] $ and makes possible the application of the formalism we discussed above. We follow the next steps.

The Young Inequality for $x,\,y,\,p_1,\,p_2>0$ and $p_1+p_2=1$ is expressed as
\begin{equation}
x\,y\leqslant p_1\,x^{1/p_1}+p_2\,y^{1/p_2}.
\end{equation}
By substituting $x=p_1^{p_1(q-1)}$ and $y=p_2^{p_2(q-1)}$ we rewrite the last equation as
\begin{equation}
\label{eq:11} p_1^{p_1(q-1)}p_2^{p_2(q-1)}\leqslant p_1^q+p_2^q\,, \qquad(q\in\mathbb{R}).
\end{equation}
Now, if we use a finite set of variables $\{x_i\}_{i=1,\,\cdots,\,W}$, Eq. (\ref{eq:11}) can be extended as follows
\begin{align}\label{eq:12}
\begin{aligned}
p_1^{p_1(q-1)}\cdots p_W^{p_W(q-1)}&\leqslant p_1^q+\cdots+p_W^q\\
\Longrightarrow\qquad \prod_{i=1}^W p_i^{(q-1)p_i}&\leqslant \sum_{i=1}^{W}p_i^q\,, \qquad(q\in\mathbb{R}).
\end{aligned}
\end{align}
From Eq. (\ref{eq:12}) one can observe two things. First, after applying the logarithmic function on both sides, we obtain the known inequality
\begin{align}\label{eq:13}
\begin{aligned}
\ln{\left(\prod_{i=1}^W p_i^{(q-1)p_i}\right)}&=(1-q)S_{\BG} \leqslant \ln{\left(\sum_{i=1}^{W}p_i^q\right)} \quad\Longrightarrow\quad
S_{q\geqslant 1}^{\Re}\leqslant S_{\BG}\leqslant S_{q\leqslant 1}^{\Re}.
\end{aligned}
\end{align}
Second, we recall that a convex function $\phi$ satisfies the relation
\begin{equation}
\label{eq:14} \phi\left(\sum_{i=1}^{W}p_i g(x_i)\right)\leqslant\sum_{i=1}^{W}p_i \phi(g(x_i)), \qquad \left(\sum_{i=1}^W p_i=1\right).
\end{equation}
For $\phi(x)=\exp{(x)}$, $g(x_i)=\ln{(x_i)}$ and $x_i=p_i^{q-1}$ we obtain
\begin{equation}
\label{eqv2:15} e^{(1-q)S_{\BG}}=\prod_{i=1}^W p_i^{(q-1)p_i}=e^{\av{\ln(p_i^{q-1})}}\leqslant \av{e^{\ln(p_i^{q-1})}}=\sum_{i=1}^{W}p_i^q.
\end{equation}
Eq. (\ref{eqv2:15}), because of Eq. (\ref{eq:12}), is always valid. Accordingly, the expression $\exp{[(1-q)S_{\BG}]}$ is a convex function for
all $q$'s and thus its second derivative is positive. Then we get the relation
\begin{equation}
\label{eqv2:16} (1-q)^2\left(\frac{\partial S_{\BG}(p)}{\partial p}\right)^2 + (1-q)\frac{\partial^2 S_{\BG}(p)}{\partial p^2}>0,
\qquad(q\in\mathbb{R}).
\end{equation}
Eq. (\ref{eqv2:16}) is interesting for $q<1$ since the first additive term is positive and the second negative. For $q>1$ the validity of
(\ref{eqv2:16}) is trivial. By applying in Eq. (\ref{eq:12}) the exponential function and after making some manipulations we can easily show
that
\begin{align}\label{eq:19}
\begin{aligned}
\left| \prod_{i=1}^{W} p_i^{(q-1)p_i}-\prod_{i=1}^{W}{p'}_{i}^{(q-1){p'}_i}\right| &\leqslant
\left| \sum \limits_{i=1}^{W}p_{i}^{q}-\sum\limits_{i=1}^{W}{p'}_{i}^{q}\right|,  \\
\Longrightarrow\qquad \left| S_{q}^{\NeG}(p) -S_{q}^{\NeG}(p') \right| &\leqslant \left| S_{q}^{\Ts}(p) -S_{q}^{\Ts}(p') \right|.
\end{aligned}
\end{align}
Taking into account Eq. (\ref{eq:7}) we extend Eq. (\ref{eq:19}) to
\begin{equation}\label{eq:20}
\left| \frac{S_{q}^{\NeG}(p) -S_{q}^{\NeG}(p') }{S_{q,\max }^{\NeG}}\right| \leqslant \left| \frac{S_{q}^{\Ts}(p) -S_{q}^{\Ts}(p')
}{S_{q,\max}^{\Ts}}\right|  .
\end{equation}
In Ref. \cite{Abe2002} it has been proved that the Tsallis entropy is Lesche stable for $q>0$. Thus from Eq. (\ref{eq:20}) we obtain that the
nonextensive Gaussian entropy is also Lesche stable for $q>0$. Now, we still have to check this criterion for negative values of $q$, since for
$q<0$ the exponential state functional does not lead to any singularity, as in the case of the trace-form entropies (see Eq. (\ref{eq:6})). But
this is already done, because the proof steps in Ref. \cite{Abe2002} can be extended without need of modifications into two regions, one for
$q<1$ and one for $q>1$. Thus we finally obtain
\begin{equation}
\left| \frac{S_{q}^{\NeG}(p) -S_{q}^{\NeG}(p') }{S_{q,\max }^{\NeG}}\right| \leqslant \left| \frac{S_{q}^{\Ts}(p) -S_{q}^{\Ts}(p')
}{S_{q,\max}^{\Ts}}\right| \leqslant
\begin{cases}
{\left( {\left\| p-p'\right\| }_{1}\right) }^{q} & \left( q<1\right)\\
q {\left\| p-p'\right\| }_{1} & \left( q>1\right)  \\
\end{cases} ,
\end{equation}
in the limit $W\rightarrow \infty $. Therefore, taking ${\|p-p'\|}_{1}<\delta _{\varepsilon }\leqslant \varepsilon ^{1/q}$ for $q<1$ or
${\|p-p'\|}_{1}<\delta _{\varepsilon }\leqslant \varepsilon /q$ for $q>1$ we see that the condition (\ref{eq:9}) is satisfied. Consequently the
nonextensive Gaussian entropy is Lesche stable for all $q\in \mathbb{R}$.

\section{Thermodynamic stability}
\label{sec3}

It is well known that BG entropy is composable and additive. From the entropy maximum principle and the additivity of $S_{\BG}$ one can derive
the thermodynamic stability condition, which is expressed as follows:
\begin{equation}
\label{eq:22} \frac{\partial ^{2}S_{\BG}( U) }{\partial U^{2}}\leqslant 0.
\end{equation}
Let us consider an isolated system composed of two identical subsystems in equilibrium. The total entropy would be $S_{\BG}( U,U) =2S_{\BG}( U)
$. Transferring now an amount of energy $\Delta  U$ from the one subsystem to the other the total entropy changes as $S_{\BG}( U+\Delta
U,U-\Delta U)=S_{\BG}( U+\Delta U)+S_{\BG}( U-\Delta U) $. According to the maximum entropy principle the final value of the entropy can not be
larger than the initial one, consequently
\begin{equation}
\label{eq:23} 2S_{\BG}( U) \geqslant S_{\BG}( U+\Delta  U) +S_{\BG}( U-\Delta  U).
\end{equation}
This is the thermodynamic stability condition for the BG entropy. In the limit $\Delta  U\rightarrow 0$ Eq. (\ref{eq:23}) tends to Eq.
(\ref{eq:22}). However, in the case of nonextensive entropies the concavity condition (\ref{eq:22}) does not correspond to thermodynamic
stability. Considering again the maximum entropy principle ($S(U,U)\geqslant S(U+\Delta U,U-\Delta U)$ for any entropy $S$) as in Eq.
\refe{eq:23} and taking this time into account the pseudo-additivity Eq. \refe{eq:8}, the TSC of $S_{q}^{\NeG}$ is written as \cite{Wada2004}
\begin{align}\label{eq:24}
\begin{aligned}
2S_{q}^{\NeG} \left( U\right) +{\left( 1-q\right) [ S_{q}^{\NeG} \left(U\right) ] }^{2 } &\geqslant
S_{q}^{\NeG} \left( U+\Delta  U\right) +S_{q}^{\NeG} \left( U-\Delta U\right)\\
& +\left( 1-q\right) S_{q}^{\NeG} \left( U+\Delta  U\right) S_{q}^{\NeG} \left( U-\Delta  U\right)  .
\end{aligned}
\end{align}
In the limit $\Delta  U\rightarrow 0$ this condition can be rewritten in the differential form
\begin{equation}
\label{eq:25} \frac{\partial ^{2}S_{q}^{\NeG}( U) }{\partial U^{2}} + \left( 1-q\right) \left\{ S_{q}^{\NeG}( U) \frac{\partial
^{2}S_{q}^{\NeG}( U) }{\partial U^{2}} - {\left(\frac{\partial S_{q}^{\NeG}( U) }{\partial U}\right) }^{2} \right\} \leqslant 0 .
\end{equation}
The TSC (\ref{eq:25}) is valid for every entropy, whose composability rule is given by Eq. (\ref{eq:8}). Replacing Eq. (\ref{eq:4}) in Eq.
(\ref{eq:25}) we obtain for the nonextensive Gaussian entropy
\begin{equation}
\label{eq:26} e^{2\left( 1-q\right) S_{\BG}( U) }\frac{\partial ^{2}S_{\BG}(U) }{\partial U^{2}} \leqslant 0.
\end{equation}
Since the exponential term is always positive and the second derivative of the BG entropy is always negative, Eq. (\ref{eq:26}) is satisfied for
every $q$. Accordingly, $S_q^G$ is thermodynamically stable for all values of $q\in \mathbb{R}$. A different way to obtain the TSC specific for
the nonextensive Gaussian entropy is the condition Eq. (\ref{eq:22}) itself, since the BG entropy can be written as
\begin{equation}
\label{eq:27} S_{\BG}=\frac{1}{1-q}\ln [ 1+\left( 1-q\right) S_{q}^{\NeG}].
\end{equation}
Then from both Eqs. (\ref{eq:22}) and (\ref{eq:27}) we obtain
\begin{align}
\label{eq:28}
\begin{aligned}
& X\times \left\{ \frac{\partial ^{2}S_{q}^{\NeG}( U) }{\partial U^{2}} + \left( 1-q\right) \left[ S_{q}^{\NeG}( U) \frac{\partial
^{2}S_{q}^{\NeG}( U) }{\partial U^{2}} - {\left( \frac{\partial S_{q}^{\NeG}( U) }{\partial U}\right)}^{2} \right] \right\}
\leqslant 0, \\
& X={\left[ 1+\left( 1-q\right)  S_{q}^{\NeG}( U) \right] }^{-2}.
\end{aligned}
\end{align}
The first multiplicative term $X$ is always positive and accordingly Eq. (\ref{eq:28}) reduces to Eq. (\ref{eq:25}).

\section{MaxEnt Distributions}
\label{sec4}

In this section we want to explore the structure of the distributions which maximize $S_{q}^{\NeG}$ under appropriate constraints. For the
internal energy $U$ the constraint we shall use is the so called escort mean value \cite{AbeOkamoto2001} expressed as
\begin{equation}
\label{eq:29} {\left\langle  U\right\rangle  }_{q} = \frac{\sum_{\ell=1}^{W}p_{\ell}^{q}U_{\ell}}{\sum_{\ell=1}^{W}p_{\ell}^{q}} ,
\end{equation}
where $U _{\ell}$ describes the energy levels of the system under consideration. We consider now the generalized canonical ensemble described by
the entropy $S_{q}^{\NeG}$ under the energy constraint (\ref{eq:29}) and the normalization constraint $\sum_{\ell=1}^{W}p_{\ell}=1$. To derive
the maximum-entropy (MaxEnt) distribution of this ensemble, we introduce the Lagrange multipliers $\alpha $ and $\beta $ and the function
\begin{equation}
I( \left\{ p_{\ell}\right\} ) =S_{q}^{\NeG}( \left\{ p_{\ell}\right\} ) +\alpha\left( 1-\sum \limits_{\ell=1}^{W}p_{\ell}\right) +\beta \left(
\av{U}_q-\frac{\sum_{\ell=1}^{W}p_{\ell}^{q}U _{\ell}}{\sum_{\ell=1}^{W}p_{\ell}^{q}}\right).
\end{equation}
We require that the variation $\delta  I$ vanishes for all perturbations $\delta  p_{i}$ of the MaxEnt distribution, accordingly
\begin{equation}
\label{eq:31} \frac{\delta I( \left\{ p_{\ell}\right\} ) }{\delta p_{i}} =-e^{\left(q-1\right) \av{\ln(p_{\ell})}}( 1+\ln(p_{i})) -\beta  q
\frac{p_{i}^{q-1}}{\sum_{\ell=1}^{W}p_{\ell}^{q}} \rbr{U_{i}-\av{U}_{q}} -\alpha \overset{!}{=}0
\end{equation}
For $q\rightarrow 1$ (\ref{eq:31}) reduces to the usual condition for Shannon's maximizer. We multiply now both sides by $p_{i}$ and sum over
$i$. Taking into account the normalization condition we obtain
\begin{equation}
\label{eq:32} \alpha =-\rbr{1+\av{\ln(p_{\ell})}} e^{( q-1) \av{\ln(p_{\ell})}}
\end{equation}
Replacing result (\ref{eq:32}) in (\ref{eq:31}) we get
\begin{equation}
\alpha \frac{\av{\ln(p_{\ell})}}{1+\av{\ln(p_{\ell})}} =\frac{\alpha }{1+\av{\ln(p_{\ell})}}\ln(p_{i}) -\frac{\beta  q
p_{i}^{q-1}}{\sum_{\ell=1}^{W}p_{\ell}^{q}} \rbr{U_{i}-\av{U}_{q}}.
\end{equation}
With the following three substitutions
\begin{equation}
\kappa =\frac{1}{\av{\ln(p_{\ell})}}, \qquad E_{i}(q) =\rbr{U_{i}-\av{U}_{q}}, \qquad \beta_{q}=\beta\frac{q}{\sum_{\ell=1}^W p_{\ell}^q}
\end{equation}
we receive the final form
\begin{equation}
1=\kappa  \ln(p_{i})+p_{i}^{q-1}\beta _{q}\kappa\,e^{(1-q)/\kappa} E_{i}( q)  .
\end{equation}
A possible way to solve this equation is by using the Lambert $W$--function \cite{Corless1996}, given as
\begin{equation}
\label{eq:36} W( x) e^{W( x) }=x .
\end{equation}
Accordingly, the distribution acquires the following structure
\begin{equation}
\label{prob} p_{i}=\frac{1}{Z}\,\exp{
                                                \left(
                                                            \frac{W\left[-(1-q)\beta_{q}E_{i}(q)\right]}{1-q}
                                                \right)}.
\end{equation}
with $Z$ given as
\begin{equation}
\label{partfunZ} Z=\sum_{\ell=1}^W \exp{
                                                \left(
                                                            \frac{W\left[-(1-q)\beta_{q}E_{\ell}(q)\right]}{1-q}
                                                \right)}.
\end{equation}
Here we can define a new generalized exponential function
\begin{equation}\label{eq:38}
e_{q}^{\mathcal{L}}( x):= \left\{
\begin{array}{ll}
 \exp{\left(\frac{W[(1-q)x]}{1-q}\right)} & , (1-q) x>0  \\
0 & , (1-q) x\leqslant 0
\end{array}
\right. ,
\end{equation}
which we call $q_{\mathcal{L}}$-exponential. Eq. (\ref{eq:38}) is symmetric with respect to $1-q$ and because of Eq. (\ref{eq:36}) it can be
written in the following different ways:
\begin{equation}\label{eq:39}
e_{q }^{\mathcal{L}}( x) = \left[ \frac{W[ (1-q)  x] }{(1-q) x}\right]^{-\frac{1}{1-q}} =\left[ \frac{W[ -(1-q)  x] }{-(1-q)
x}\right]^{\frac{1}{1-q}} =e^{\frac{W[(1-q) x]}{(1-q)}} =e^{\frac{W[-(1-q) x]}{-(1-q)}}.
\end{equation}
For $q =1$ the generalized exponential function in (\ref{eq:38}) and (\ref{eq:39}) tends to the ordinary one.

As we can see the form of the probability distribution (\ref{prob}) is not very familiar and clearly distinguishes from the R\'enyi/Tsallis
ones. However, in the asymptotic limit, which is, for $E_i(q)\gg 1/[\beta_{q}(q-1)]$, Eq. (\ref{prob}) tends to a power-law distribution
function
\begin{equation}
\label{eq:41} p_i \propto
                \left[
                E_{i}(q)
                \right]^{\frac{1}{1-q}}
= \Big(
    U_{i}-\langle U\rangle_{q}
    \Big)^{\frac{1}{1-q}},
\end{equation}
same as the R\'enyi/Tsallis maximum entropy distribution $p_i$.

\section{Connection to thermodynamics}\label{sec5}

In Refs. \cite{TsallisMendesPlastino} and \cite{LenziMendesSilva} it has been shown that the entire Legendre structure of thermodynamics is
$q$-invariant with regard to Tsallis and R\'enyi entropy respectively. In this section we explore whether the Legendre structure is also
invariant with respect to NeG entropy. Using Eq. (\ref{eq:39}) we can express the Lambert function in Eq. (\ref{prob}) in the following two ways
\begin{equation}\label{eq:42}
W\left[(q-1)\beta_{q}E_i(q)\right]= \left\{
\begin{array}{l}
 (p_i Z)^{q-1} (q-1)\beta_{q}E_i(q)  \\
(1-q)\ln{(p_i Z)}
\end{array}
\right..
\end{equation}
Accordingly, we have
\begin{equation}\label{eq:43}
(p_i Z)^{q-1}\beta_{q}E_i(q)=-\ln{(p_i Z)}.
\end{equation}
By multiplying Eq. (\ref{eq:43}) with $p_i$ and taking the sum over all $i$'s the left hand side of Eq. (\ref{eq:43}) vanishes because of the
constraint $\sum_{i=1}^{W}p_i^q U_i=\av{U}_q\sum_{i=1}^{W}p_i^q$. Then we obtain
\begin{equation}\label{eq:44}
Z=e^{-\av{\ln(p_i})}=\prod_{i=1}^W p_i^{-p_i}.
\end{equation}
Consequently we can express the entropy $S_q^{\NeG}$ in dependence on $Z$ as
\begin{equation}\label{eq:45}
S_q^{\NeG}=\ln_q(Z).
\end{equation}
With the introduction of a temperature $1/T=\partial S_q^{\NeG}/\partial \av{U}_q$ \cite{PlastinoPlastino1997}, where $T$ is connected with the
Lagrange multiplier $\beta$ as $\beta:=1/T$, and after defining the partition function $\tilde{Z}$ as
\begin{equation}\label{eq:46}
\ln_q(\tilde{Z}):=\ln_q(Z)-\beta\av{U}_q,
\end{equation}
one can show that the escort mean energy $\av{U}_q$ can be expressed as
\begin{equation}\label{Uq}
\av{U}_q:=-\frac{\partial}{\partial\beta}\ln_q(\tilde{Z}).
\end{equation}
Then, the free energy $F_q$, which is defined as
\begin{equation}\label{eq:47}
F_q:=\av{U}_q-T\,S_q^{\NeG}=\av{U}_q-\frac{1}{\beta}S_q^{\NeG},
\end{equation}
can be written as
\begin{equation}\label{eq:48}
F_q=-\frac{1}{\beta}\ln_q(\tilde{Z}),
\end{equation}
for the maximum entropy distribution (\ref{prob}). We can also verify that
\begin{equation}\label{eq:51}
C_q:=T\frac{\partial S_q^{\NeG}}{\partial\beta} =\frac{\partial \av{U}_q}{\partial\beta} =-\,T\frac{\partial^2F_q}{\partial T^2},
\end{equation}
where $C_q$ is the generalized specific heat. In other words, the NeG entropy under the constraint of the internal energy (\ref{eq:29}) and the
normalization constraint, preserves the Legendre structure of thermodynamics.

Next we shall present the relation between the generalized temperature and specific heat with the ordinary ones. In the BG case the temperature
and the specific heat are given by
\begin{equation}\label{eq:52}
\frac{1}{T}=\frac{\partial S_{\BG}}{\partial U}, \qquad \frac{1}{C}=\frac{\partial T}{\partial U}=-T^2\frac{\partial^2 S_{\BG}}{\partial U^2},
\qquad (U=\av{U}_1).
\end{equation}
By replacing the BG entropy in Eq. (\ref{eq:52}) with Eq. (\ref{eq:27}) we obtain
\begin{align}\label{eq:53}
\begin{aligned}
T&= \left\{1+(1-q)S_q^{\NeG}\right\}T_q \qquad\mathrm{with}\qquad
\frac{1}{T_q}:=\frac{\partial S_q^{\NeG}}{\partial U},\\
\frac{1}{C}&=\frac{1}{C_q}+(1-q)\left(1+\frac{S_q^{\NeG}}{C_q}\right) \qquad\mathrm{with}\qquad \frac{1}{C_q}:=-T_q^{2}\frac{\partial^2
S_q^{\NeG}}{\partial U^2}
\end{aligned}
\end{align}
These two expressions for $T$ and $C$ are the same with those derived for the Tsallis entropy. Wada in Ref. \cite{Wada2004} computed the
relations in (\ref{eq:53}) from the composition rule (\ref{eq:8}). Accordingly, they are valid for every entropy that satisfies Eq.
(\ref{eq:8}).

Finally, in Eq. \refe{eq:6} we showed, that by using the $q$-logarithm \refe{eq:7} there two possible ways to generalize the BG entropy. Here we
explore the essence of this result and show that for an arbitrary generalized logarithm the BG entropy can be generalized actually in three
different ways.

Therefore, we consider an isolated system composed by $N$ independent particles, with their energy levels characterized by the occupation
numbers $n_1,\,n_2$,\\$\cdots,\,n_W$ and their respective probabilities $p_1,\,p_2,\,\cdots,\,p_W$. Then the number of all possible
configurations of the particles is given by the multinomial coefficient $M$:
\begin{equation}
M:=\left[\frac{N!}{(n_{1}) !( n_{2}) !\cdots ( n_{W}) !}\right]
    =\left[\frac{N!}{(N p_{1}) !( N p_{2}) !\cdots ( N p_{W}) !}\right].
\end{equation}
In further we introduce the quantity $\mathcal{X}:=M^{1/N}$. For $N\rightarrow \infty $ and taking into account the relation
$\lim_{N\rightarrow\infty}N!\approx {(\frac{N}{e})}^{N}$, we can easily show that
\begin{equation}\label{eq:w1}
\mathcal{X} =\prod \limits_{i=1}^{W}p_{i}^{-p_{i}} =e^{\av{\ln{\rbr{1/p_{i}}}}}.
\end{equation}
Now, the BG entropy is defined in thermal equilibrium as the application of the logarithmic function on Eq. (\ref{eq:w1}):
\begin{equation}\label{eq:w2}
S_{\BG}:=\ln{ ( \mathcal{X})} =\ln{\rbr{\prod_{i=1}^{W}p_i^{-p_i}}} =\av{\ln{\rbr{1/p_{i}}}} =-\av{\ln{\rbr{p_{i}}}}.
\end{equation}
Although all expressions in Eq. \refe{eq:w2} are equal, it is obvious that the replacement of a generalized logarithmic function $L_{\vec{q}}$
with
\begin{equation}
\lim_{\vec{q}\rightarrow\vec{q}_{0}}L_{\vec{q}}(a) =\ln{\rbr{a}} , \qquad(a>0),
\end{equation}
and a set of parameters $\vec{q}:=\{q_{i}\}_{i=1,\,\cdots,\,m}$, leads to different generalized entropy structures. These are the following:
\begin{equation}\label{eq:w3}
S_{\vec{q}}^{(1)}=L_{\vec{q}}\rbr{\prod_{i=1}^{W}p_i^{-p_i}}, \quad S_{\vec{q}}^{(2)}=\av{L_{\vec{q}}\rbr{1/p_{i}}}, \quad
S_{\vec{q}}^{(3)}=-\av{L_{\vec{q}}(p_{i})}.
\end{equation}
There are three things to notice. First, the R\'enyi definition does not correspond to any of the three entropy generalizations in Eq.
(\ref{eq:w3}). Second, $S_{\vec{q}}^{(2)}$ and $S_{\vec{q}}^{(3)}$ have the same structure . The small differences between them can be refered
to a transformation with respect to $\vec{q}$ ($\vec{q}^{(2)}\rightarrow f(\vec{q}^{(3)})$). Thus they represent actually the same quantity.
Third, the maximization of $S_{\vec{q}}^{(1)}$ under consideration of the constraint \refe{eq:29} leads always to Lambert exponential
distributions, independent from the choice of the $\vec{q}$-logarithm, because of the following relation
\begin{equation}\label{eq:w4}
\frac{\partial S_{\vec{q}}^{(1)}}{\partial p}= \frac{\partial L_{\vec{q}}\rbr{S_{\BG}}}{\partial S_{\BG}}\, \frac{\partial S_{\BG}}{\partial p}.
\end{equation}
Using the one-parametric generalized logarithm \refe{eq:7} we identify $S_q^{(1)}=S_q^{\NeG}$, $S_q^{(2)}=S_q^{\Ts}$ and $S_q^{(3)}$ is the
$S_{2-q}^{\Ts}$ transformed Tsallis entropy ($q\rightarrow2-q$).

\section{Conclusions}

We have studied some statistical properties of the nonextensive Gaussian entropy (\ref{eq:4}). $S_q^G$ is Lesche stable (or experimentally
robust) for all values of $q\in\mathbb{R}$. We have shown that the Lesche stability of $S_{q}^{\NeG}$ is a consequence of the Lesche stability
of the Tsallis entropy. We found the same thermodynamical stability condition as in the case of Tsallis entropy with regard to the ordinary
internal energy, using the concavity condition of the Boltzmann--Gibbs entropy, since $S_{q}^{\NeG}$ can be expressed as functional of the
entire $S_{\BG}$. The condition is satisfied for all values of $q\in\mathbb{R}$. We derived the distribution that maximizes the nonextensive
Gaussian entropy. This is based on the Lambert $W$--function. A new generalized $q_{\mathcal{L}}$-exponential function is defined. For $q=1$ it
returns to the ordinary one. In the thermodynamic limit it tends to a pure power-law function. The connection of $S_q^G$ to thermodynamics is
presented. We showed that the Legendre structure is preserved through a convenient definition of a generalized partition function and the
relation between the temperature and specific heat with the generalized ones is the same as in the case of Tsallis entropy. Finally, we
demonstrated that by replacing the ordinary logarithm in the equilibrium Boltzmann--Gibbs entropy with a generalized one, we obtain three
possible entropy structures, in which one can identify the nonextensive Gaussian entropy and the Tsallis entropy. The R\'enyi entropy, since it
is not based on the concept of a generalized logarithmic function, does not belong to any of these three cases.

\section*{Acknowledgments}
Special thanks to Prof. S. Abe and Dr. A. Provata for discussions and important comments.


\end{document}